\pdfoutput=1
\documentclass[12pt]{article}
\usepackage[utf8]{inputenc} 
\usepackage[T1]{fontenc}
\usepackage{hyperref}
\usepackage{url}      
\usepackage{setspace}
\usepackage[pdftex]{graphicx}
\usepackage{dcolumn}
\usepackage[margin=1in]{geometry}
\usepackage{amsmath}
\usepackage{graphicx}
\usepackage{float}
\usepackage{tabularx}
\usepackage[sort, numbers]{natbib}
\usepackage[mathscr]{eucal}
\usepackage{multirow}
\usepackage{booktabs}
\usepackage{xcolor}

\usepackage{caption}
\usepackage{setspace}

\providecommand{\keywords}[1]
{
  \small	
  \textbf{\textit{Keywords---}} #1
}

\begin{document}

\title{Slow-roll Hilltop Inflation in $f(\phi,T)$ gravity}

\author{ Biswajit Deb  \footnote{Electronic Address:biswajitdeb55@gmail.com} , Atri Deshamukhya  \footnote{Electronic Address: atri.deshamukhya@aus.ac.in}\\
Department of Physics, Assam University, Silchar, India }

\date{}
\maketitle

\begin{abstract}
Over the last four decades, a number of modified gravity theories have been proposed to study cosmological phenomena as they can provide solutions for some of the shortcomings of Einstein's gravity in explaining early and late time accelerations of the observed Universe, the existence of dark matter, singularities at center of Black holes etc. The theoretical and observational challenges faced by the $\Lambda$CDM model also point towards the necessity for looking beyond General Relativity. In this direction, recently $f(\phi, T)$ gravity has been proposed in literature where the non-minimal coupling of the scalar field $\phi$ with the trace of energy-momentum tensor $T$ has been introduced in the Einstein-Hilbert action. Considering the Hilltop potential, we have studied the slow-roll inflation in the framework of $f(\phi, T)$ gravity. It is found that Hilltop inflationary models in $f(\phi, T)$ gravity are viable when seen in the light of latest Planck data.

\end{abstract} \hspace{8pt}


\section{Introduction}	
Inflation is one of the cornerstones of the standard model of cosmology. Since its inception in the early 80's by Alan Guth to solve the horizon and flatness problem associated with the Big Bang model \cite{r1}, the inflationary hypothesis could explain the latest observations to a great extent. For example, it can generate the primordial density fluctuations which later on manifest into the large-scale structures of the Universe \cite{r2,r3,r4,r5,r6} supported by the observational results from COBE, WMAP and Planck \cite{r7,r8,r9}. The simplest model of inflation is achieved by a canonical scalar field called inflaton which rolls down slowly on its potential giving rise to a de-sitter expansion \cite{r1}. Once the inflation ends, the Universe enters into radiation radiation-dominated era through a reheating phase \cite{r10,r11} and particle production begins via leptogenesis, baryogenesis, and nucleosynthesis at different energy scales \cite{r12,r13,r14,r15}. This provides a simple yet concrete base theory for the early evolution of the Universe (For comprehensive review \cite{r16,r17,r18} can be seen). However, non-canonical and vector fields are also used in literature as a driving source of inflation \cite{r19,r20,r21,r22,r23}.
\newpage
The $\Lambda$CDM model based on General Relativity (GR) is so far the best fit theoretical model for describing the Universe \cite{r24}. However, it is yet to explain the origin and smallness of the cosmological constant $\Lambda$ amongst a few other issues \cite{r25}. Einstein's GR is successful at the small scales but it has limitations at the large scales e.g., it can't explain the dark sector as well as the current accelerated expansion of the Universe \cite{r26,r27}. GR can't explain cosmic coincidence as well \cite{r28}. These loopholes in GR strongly motivated the cosmologist to look for modifications/alternative theories to Einstein's gravity \cite{r29}. (For detail review see \cite{r30,r31}) \\
\\
The earliest modification in the gravitational Lagrangian was done by Weyl and Eddington on the basis of curiosity only \cite{r32,r33}. Later on in the 1980's, Starobinsky came out with the $R^2$ inflation model whose result shocked the world \cite{r34}. The Starobinsky model, broadly speaking the $f(R)$ gravity model, is not only capable of describing inflation, but it can also explain the late time expansion without the requirement of an exotic energy component in the matter sector \cite{r35,r36,r37,r38}. In fact, $R^2$ inflation is the best-fit inflation model as per the latest Planck result \cite{r39}. Modification in the GR action can be done either in the geometry part or in the matter part or in both by adding higher order scalar terms maintaining the Lorentz invariance of the action \cite{r29}. In the past four decades, a bunch of modified theories of gravity emerged in the literature viz. $f(G)$, $f(G,T)$, $f(Q)$, $f(R,T)$, etc. \cite{r40,r41,r42,r43} which are quite successful in studying astrophysical objects and cosmological phenomena \cite{r44,r45,r46,r47,r48,r49,r50,r51,r52,r53}. Here G, T, Q stands for Gauss-Bonnet scalar, trace of EM tensor and non-metricity respectively.\\
\\
Of many such modified theories, $f(\phi)T$ gravity has been proposed recently by Zhang et al. where the non-minimal coupling of the scalar field $\phi$ with the trace of EM tensor $T$ has been added to the gravitational Lagrangian \cite{r54}. This kind of addition of coupling terms is generally motivated by the quantum gravity framework because it provides a more comprehensive description of gravity unified with other fields which pave the way for the exploration of new physics beyond the scope of general relativity \cite{r55,r56,r57,r58,r59}. The charm of $f(\phi)T$ gravity is that at the end of inflation when the field is decayed to radiation, Einstein's gravity is recovered naturally \cite{r54}. Now technically, $f(\phi)T$ gravity is a simple extension of $f(R,T)$ gravity with a coupling term. Since $f(R,T)$ gravity presents interesting results for inflation models \cite{r60,r61,r62,r63,r64}, $f(\phi)T$ gravity holds a scope too. Zhang et al. studied the slow-roll inflation with $f(\phi)T = \sqrt{\kappa}\phi T$ and reported that Chaotic, Natural, and Starobinsky potentials are in better agreement with data \cite{r54}. Besides this, Ashmita et al. studied other potentials with higher-order coupling terms \cite{r65} and Herrera et al. studied the reheating mechanism in $f(\phi)T$ gravity \cite{r66}. These affirmative results initiate the curiosity to investigate leading inflationary models within the $f(\phi)T$ gravity setup.  \\
\\
Now with the advancement of observational techniques, tighter bounds on the spectral tilts and scalar-to-tensor ratio have been imposed \cite{r39}. This results in the rejection of several inflationary potentials \cite{r39,r67}. Planck 2018 reported that the quartic hilltop potential falls within the allowed range at the super Planckian scales \cite{r39}. But, super Planckian scales are problematic from theoretical perspectives \cite{r68}. Hence efforts are being made to lower the inflation scale for quartic hilltop potential in $f(R,T)$ gravity \cite{r64}. Generally, hilltop potentials are a class of small field models that give red tilted spectral index in GR at super Planckian scales \cite{r69}. In this work, we will investigate whether hilltop potentials can meet observational bounds within the framework of $f(\phi,T)$ gravity at lower energy scales. We have taken a more generalised form of the coupling term as $f(\phi,T)$ instead of $f(\phi)T$ which allows broader scope of work. \\
\\
The paper has been organized as follows. In section 2, we review  the formalism of slow-roll inflation in the framework of $f(\phi,T)$ gravity. In section 3, we present results for hilltop potentials and finally, in section 4 we conclude. Throughout the paper, we use natural units $\hbar=G=c=1$ and follow (-,+,+,+) convention for the metric tensor.

\section{Slow-roll inflation in $f(\phi,T)$ gravity}
Form of the action in $f(\phi,T)$ gravity is 
\begin{equation}
    S= \int d^4x \sqrt{-g} \left[\frac{R}{2\kappa} + f(\phi,T) + L_m \right]
    \label{1}
\end{equation}
where $R= g^{\alpha\beta}R_{\alpha\beta}$ is the scalar curvature, $f(\phi,T)$ is an arbitrary function of scalar field $\phi$ non minimally coupled with the trace $T$ of the energy-momentum tensor, $L_m$ is the Lagrangian of the matter sector and $\kappa=8 \pi G = \frac{1}{{M_{Pl}}^2}$ and, $M_{Pl}$ is the reduced Planck mass. \\
\\
To initiate and drive inflation in the early universe, homogeneous scalar fields are required. The canonical scalar field $\phi=\phi(t)$ also known as inflaton is introduced by the Lagrangian of the form
\begin{equation}
     L_m = -\frac{1}{2}g^{\alpha\beta}\partial_{\alpha}\phi \partial_{\beta}\phi - V(\phi) = \frac{\dot \phi^2}{2} - V(\phi)
     \label{2}
\end{equation}
where $V(\phi)$ is the potential of the scalar field. Then the energy-momentum tensor $T_{\alpha\beta}$ can be calculated form the matter Lagrangian as
\begin{equation}
    T_{\alpha\beta} = g_{\alpha\beta}L_m - 2 \frac{\delta L_m}{\delta g^{\alpha\beta}} = \partial_{\alpha}\phi \partial_{\beta}\phi + g_{\alpha\beta} \left [ \frac{\dot \phi^2}{2} - V(\phi) \right ]
    \label{3}
\end{equation}
Assuming the universe filled with a perfect fluid, the energy density $\rho$, pressure $p$ components of $T_{\alpha\beta}$ and the trace $T$ are obtained as
\begin{equation}
    T_{00} = \frac{\dot\phi^2}{2} + V(\phi) = \rho
    \label{4}
\end{equation}
\begin{equation}
    T_{ij} = \left[\frac{\dot\phi^2 }{2} - V(\phi)\right]g_{ij} = p g_{ij}
    \label{5}
\end{equation}
\begin{equation}
    T = g^{\alpha\beta}T_{\alpha\beta} = \dot\phi^2 - V(\phi)
    \label{6}
\end{equation}
On metric variation of the action Eq. \ref{1}, the modified field equations are obtained as
\begin{equation}
    R_{\alpha\beta} - \frac{1}{2}g_{\alpha\beta}R = \kappa \left[ T_{\alpha\beta} + g_{\alpha\beta} f(\phi,T) - 2 f_T(\phi,T)(T_{\alpha\beta} + \Theta_{\alpha\beta}) \right]
    \label{7}
\end{equation}
where $\Theta_{\alpha\beta}$ is
\begin{equation}
    \Theta_{\alpha\beta} = g^{\zeta\nu} \frac{\delta T_{\zeta\nu}}{\delta g^{\alpha\beta}} = -2T_{\alpha\beta} + g_{\alpha\beta}L_m - 2 g^{\zeta\nu} \frac{\delta^2 L_m}{\delta g^{\alpha\beta} \delta g^{\zeta\nu}} 
    \label{8}
\end{equation}
This term $\Theta_{\alpha\beta}$ contains matter Lagrangian $L_m$. So, depending on the type of matter, it's form will vary.  For inflaton field $\phi$, Eq. \ref{8} becomes
\begin{equation}
    \Theta_{\alpha\beta} = - \partial_\alpha \phi \partial_\beta \phi - T_{\alpha\beta}
    \label{9}
\end{equation}
The different components of $\Theta_{\alpha\beta}$ are computed as
\begin{equation}
    \Theta_{00} = - \dot \phi^2 - T_{00}, \hspace{0.5cm} \Theta_{ij} = - T_{ij}
    \label{10}
\end{equation}
For this particular study, the form of $f(\phi,T)$ is taken as $f(\phi,T) = \sqrt{\kappa} \lambda \phi T$ where $\lambda$ is a dimensionless model parameter and we set $\kappa = 1$ for simplicity of the calculation. With this particular choice of $f(\phi,T)$, the field equations are obtained from Eq. \ref{7} as
\begin{equation}
    R_{\alpha\beta} - \frac{1}{2}g_{\alpha\beta}R = T_{\alpha\beta}^{eff}
    \label{11}
\end{equation}
where $T_{\alpha\beta}^{eff}$ is the effective energy-momentum tensor containing both the standard matter contribution and the modification induced by the $\phi T$ coupling term. It is defined as
\begin{equation}
    T_{\alpha\beta}^{eff} = T_{\alpha\beta} - 2 \lambda \phi\left (T_{\alpha\beta} - \frac{1}{2} T g_{\alpha\beta} + \Theta_{\alpha\beta} \right)
    \label{12}
\end{equation}
It is to be noted that the modified field Eqs. \ref{11} immediately reduces to Einstein field equations when $\lambda=0$. Further at the end of inflation, when the field is completely decayed to radiation, this model returns to Einstein gravity. Now from Eq. \ref{12}, the effective energy density and effective pressure can be obtained as
\begin{equation}
    T_{00}^{eff} = \rho^{eff} = \frac{\dot \phi^2}{2} (1+2 \lambda\phi) + V(\phi) (1+4\lambda\phi)
    \label{13}
\end{equation}
\begin{equation}
     T_{ij}^{eff} = p^{eff} g_{ij} = \left[ \frac{\dot \phi^2}{2} (1+2\lambda\phi) - V(\phi) (1+4\lambda\phi) \right] g_{ij}
    \label{14}
\end{equation}
To study the cosmological implications of the field equations, the Friedmann-Lemaitre-Robertson-Walkar (FLRW) metric was proposed on the basis of cosmological principle. The FLRW metric in the spherical coordinates reads as
\begin{equation}
    ds^2= - dt^2 + a(t)^2 \left[\frac{dr^2}{1-kr^2}+r^2 (d\theta^2 + \sin^2\theta d\phi^2)\right]
    \label{15}
\end{equation}
where $a(t)$ is the scale factor, $t$ is the cosmic time and $k$ is the spatial curvature with value +1, -1, and 0 which corresponds to close, open and flat universe respectively. Since the current observation shows that the present universe is spatially flat, $k=0$ is set for the calculations. Using this metric, the modified Friedmann equations are obtained as
\begin{equation}
    3H^2 = \frac{\dot \phi^2}{2} (1+2 \lambda\phi) + V(\phi) (1+4\lambda\phi)
    \label{16}
\end{equation}
\begin{equation}
    \frac{\ddot a}{a} = - \frac{1}{3} [ \dot\phi^2 (1+2\lambda\phi) - V(\phi) (1+4\lambda\phi) ]
    \label{17}
\end{equation}
where $H=\frac{\dot a}{a}$ is the Hubble parameter. Further, the modified equation of motion can be obtained from Eqs. \ref{13} and \ref{14} as
\begin{equation}
    (\ddot \phi + 3H\dot \phi ) (1+2\lambda\phi) + \lambda \dot \phi^2 + (1+4\lambda\phi) V_{,\phi} + 4\lambda V(\phi) =0
    \label{18}
\end{equation}
where $V_{,\phi} = \frac{dV}{d \phi}$. It is to be noted that under the limit $\lambda=0$,  Eqs. \ref{16}, \ref{17} and \ref{18} reduce to their standard forms in GR. The inflation need to be continued for a prolonged time so that it can overcome the horizon problem. For this the inflaton need to roll down slowly over its potential for a sufficient time. This requires imposition of slow-roll conditions which are \cite{r17}
\begin{equation}
    \dot \phi^2 \ll V(\phi), \hspace{0.5cm} \ddot \phi \ll 3H\dot\phi, \hspace{0.5cm} \dot\phi^2 \ll H\dot\phi 
    \label{19}
\end{equation}
Under these slow-roll conditions, the modified Friedmann equation \ref{16} and modified equation of motion \ref{18} becomes
\begin{equation}
    3H^2 = V(\phi) (1+4\lambda\phi)
    \label{20}
\end{equation}
\begin{equation}
     3H\dot \phi (1+2\lambda\phi) + (1+4\lambda\phi) V_{,\phi} + 4\lambda V(\phi) =0
    \label{21}
\end{equation}
Then the modified potential slow-roll parameters are derived from Eqs. \ref{20} and \ref{21} as 
\begin{equation}
    \epsilon_{V} = \frac{1}{2(1+2\lambda\phi)} \left [ 
    \frac{V_{,\phi}}{V} + \frac{4\lambda}{1+4\lambda\phi} \right]^2 \label{22}
\end{equation}
\begin{equation}
    \eta_V = \frac{1}{1+2\lambda\phi} \left[\frac{V_{,\phi\phi}}{V} + \frac{\lambda (6 + 8\lambda\phi)}{(1+2\lambda\phi)(1+4\lambda\phi)}\frac{V_{,\phi}}{V} - \frac{8\lambda^2}{(1+2\lambda\phi)(1+4\lambda\phi)} \right] 
    \label{23}
\end{equation}
It is seen that correction from the $\phi T$ term has been induced in the slow-roll parameters which will play an important role to make the model result consistent with the observation. Further when $\lambda=0$, these slow-roll parameters reduce to their standard GR forms \cite{r17}. Inflation continues till $\epsilon_V < 1, \eta_V < 1$ and stops when either of the parameter becomes one. These slow-roll parameters are pivotal to describe the inflation dynamics and its observational imprints on CMB. One can calculate the scalar spectral index $n_s$ and tensor-to-scalar ratio $r$ using the field value at horizon crossing from relations \cite{r4,r17}
\begin{equation}
    n_s =   1 - 6 \epsilon_V + 2 \eta_V 
    \label{24}
\end{equation}
\begin{equation}
    r = 16 \epsilon_V 
    \label{25}
\end{equation}
The amount of inflation is described by the number of e-folds $N$ given by
\begin{equation}
    N = \int_{t_1}^{t_2} H dt = \int_{\phi_{end}}^{\phi} \frac{H}{\dot \phi} d\phi = \int_{\phi_{end}}^{\phi} \frac{V(\phi)(1+2\lambda\phi)(1+4\lambda\phi)}{V_{,\phi}(1+4\lambda\phi) + 4\lambda V(\phi)} d\phi
    \label{26}
\end{equation}
where $\phi_{end}$ is value of the inflaton at the end of inflation and the upper limit $\phi$ is the value of inflaton at horizon crossing. Thus, knowing the form of potential $V(\phi)$ one can study any inflation model.


\section{A case study with Hilltop potentials}

Hilltop potentials were originally introduced by Lotfi Boubekeur and David H. Lyth in 2005 \cite{r70}. Here, the inflaton potential has the shape of a hill and the inflaton rolls down from the top of the hill to its minimum hence the name "hilltop potential". The general form of the hilltop potential reads as
\begin{equation}
   V(\phi)=\Lambda^4 \left[1-\left(\frac{\phi}{\mu}\right)^m+  ...\right]
   \label{37}
\end{equation}
where $\Lambda$ is inflationary scale, $\mu$ is vacuum expectation value ($vev$) of the inflaton and $m$ is a positive integer. The ellipsis stands for higher order terms which are irrelevant during inflation but needed to stabilize the potential as the inflation ends \cite{r71}. The inflation takes place in the first quadrant $(\phi \ge 0)$ mainly around $\phi=0$. In order to ensure this, the first and second derivative of the potential at $\phi=0$ must be constrained such that $V_{,\phi}=0$ and $V_{,\phi\phi} < 0$. These two conditions hold good for positive values of $\Lambda$, $\mu$ with $m > 1$ \cite{r71}. \\
\\
Different hilltop inflationary models corresponding to different values of $m$ have been studied in the literature, particularly the quadratic ($m=2$) and the quartic ($m=4$) hilltop models \cite{r72,r73,r74,r75,r76,r77}. Recently, the analytical study of quartic hilltop model infers that the model predicts $n_s$ consistent with Planck for $\mu \gg 10 M_p$ \cite{r78}. Planck reported that the quartic model is in better agreement than the quadratic model, however $\mu \gg 10 M_p$ is required \cite{r39}. A more detail study of the different cases $m = \{3/2,2,3,4\}$ shows that in GR, they are able to meet observational bounds at super Planckian scales and the cases with higher $m$ predict better result \cite{r71}.\\
\\
When the potential in Eq.\ref{37} is used in the Eqs. \ref{22} and \ref{23}, the slow-roll parameters stands as
\begin{equation}
    \epsilon_V= \frac{\left[\left(\frac{\phi}{\mu}\right)^m\{m+4\lambda\phi(1+m)\}-4\lambda\phi\right]^2}{2\phi^2 (1+2\lambda\phi)(1+4\lambda\phi)^2\left[\left(\frac{\phi}{\mu}\right)^m-1\right]^2} \label{28}
\end{equation}
\begin{equation}
    \eta_V= \frac{8\lambda^2\phi^2+\left(\frac{\phi}{\mu}\right)^m[m(m-1)+6\lambda \phi m^2+8\lambda^2\phi^2(m^2-1)]}{\phi^2 (1+2\lambda\phi)^2 (1+4\lambda\phi) \left[\left(\frac{\phi}{\mu}\right)^m-1\right] } \label{29}
\end{equation}
The scalar spectral index and tensor-to-scalar ratio are obtained as
\begin{equation}
    n_s = 1 - \frac{3\left[\left(\frac{\phi}{\mu}\right)^m\{m+4\lambda\phi(1+m)\}-4\lambda\phi\right]^2}{\phi^2 (1+2\lambda\phi)(1+4\lambda\phi)^2\left[\left(\frac{\phi}{\mu}\right)^m-1\right]^2} + \frac{16\lambda^2\phi^2+ \left(\frac{\phi}{\mu}\right)^m 2[m(m-1)+6\lambda \phi m^2+8\lambda^2\phi^2(m^2-1)]}{\phi^2 (1+2\lambda\phi)^2 (1+4\lambda\phi) \left[\left(\frac{\phi}{\mu}\right)^m-1\right] }
    \label{30}
\end{equation}
\begin{equation}
    r = \frac{8\left[\left(\frac{\phi}{\mu}\right)^m\{m+4\lambda\phi(1+m)\}-4\lambda\phi\right]^2}{\phi^2 (1+2\lambda\phi)(1+4\lambda\phi)^2\left[\left(\frac{\phi}{\mu}\right)^m-1\right]^2} \label{31}
\end{equation}
\\
Clearly, $n_s$ and $r$ depend on $\lambda$ and $\mu$ for a particular choice of m. The $\phi$ stands for the initial value of the field during horizon crossing and is calculated numerically from Eq. \ref{26}, keeping the N fixed at 50 e-folds. The value of $\phi_{end}$ will be marked either by $\epsilon_V = 1$ or $\eta_V = 1$, depending on which one becomes unity earlier. With a suitable choice of $\lambda$, it is possible to bring $n_s$ and $r$ within the desirable range. Here we will study four different cases of hilltop inflation with $m = \{3/2,2,3,4\}$ keeping $vev$ of the inflaton fixed at $\mu = 10 M_p$ and $\mu =5 M_p$ respectively.


\subsection{When $\mu = 10 M_p$}

\begin{figure}[hbt!]
        \centering
        \includegraphics[width=0.6 \textwidth]{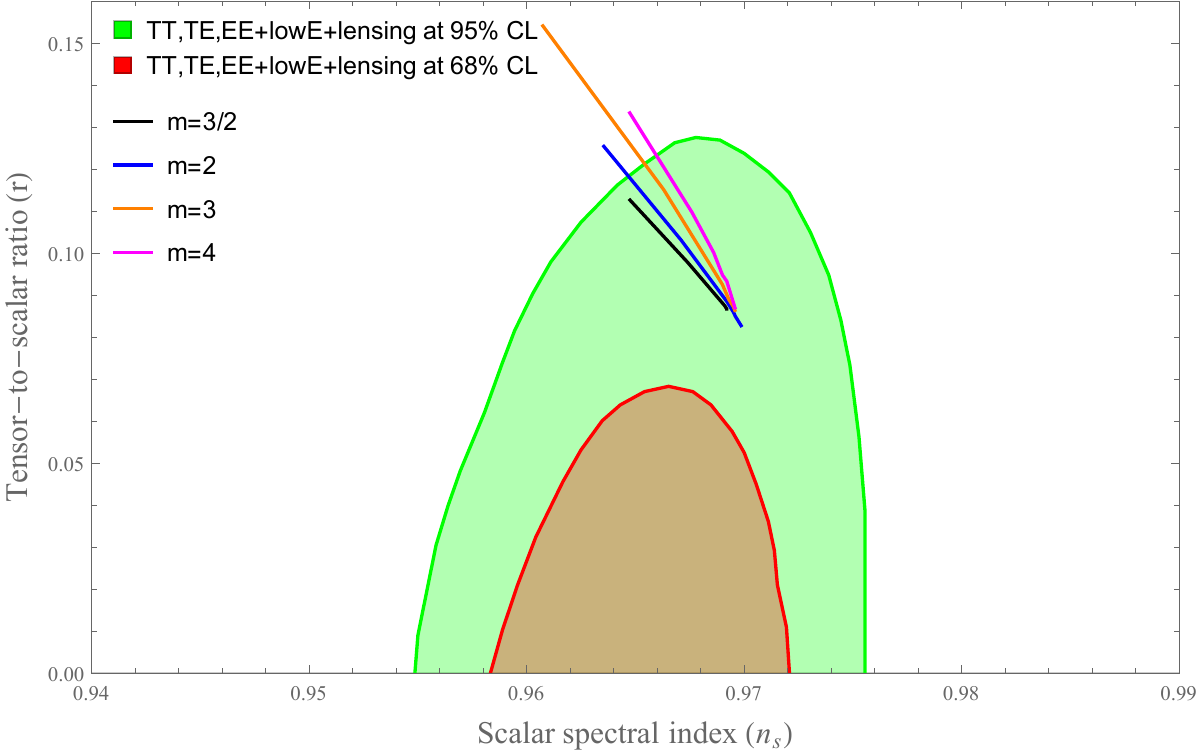}
        \caption{The $n_s-r$ plot predicted by the hilltop potentials corresponding to $m = \{3/2,2,3,4\}$ marked by black, blue, orange and pink colour respectively at $N=50$ and $\mu=10M_p$. The marginalized joint 68\% and 95\% C.L. regions for $n_s$ and $r$ at $k=0.002 Mpc^{-1}$ from Planck 2018 data are shown in red and green colour respectively \cite{r39}. \label{f1}}
\end{figure}

The $n_s-r$ plot for different cases of hilltop potential is shown in Figure \ref{f1}. The plots are obtained at $N=50$ e-folds for different values of the model parameter $\lambda$. We observe that in all cases, the scalar spectral index is red-tilted and in the range allowed by Planck 2018. With the increase in the positive values of $\lambda$, $n_s$ shifts towards higher values and the $r$ decreases gradually. Thus the correction term in the action elevates the value of $n_s$ to bring it inside the Planck range and suppresses the $r$ value. The plots for the cases $m = \{3/2,2,3,4\}$ enter into the $2\sigma$ region of Planck 2018. Further for each cases of $m$, it is found that $n_s$ tends to be constant at higher values of $\lambda$. \\
\\
We then obtained the model parameter space for which the hilltop potentials remain consistent with Planck 2018. The range of $\lambda$ is summarised in the Table \ref{t1}.
\begin{table}[hbt!]
\centering
            \begin{tabular}{lccc}
            \hline \hline
    Case   &   Range of model parameter \\ \hline
   $m=3/2$	&	 $0.7 < \lambda < 9.6$	\\
   $m=2$	&    $1.4 < \lambda < 10$	\\
   $m=3$	&    $3.7 < \lambda < 13$	\\
   $m=4$	&	 $6.8 < \lambda < 15.8$	\\
 \hline \hline
            \end{tabular} 
            \caption{Allowed parameter space for $m = \{3/2,2,3,4\}$ cases for $N=50$ at $\mu = 10 M_p$ }
            \label{t1}
            \end{table}
\\
It is seen that for a small range of $\lambda$, the potentials enter into $2\sigma$ region of Planck. The range of $\lambda$ shifts towards higher values as the values of $m$ increase. 

\subsection{When $\mu = 5 M_P$}
\begin{figure}[hbt!]
        \centering
        \includegraphics[width=0.6\textwidth]{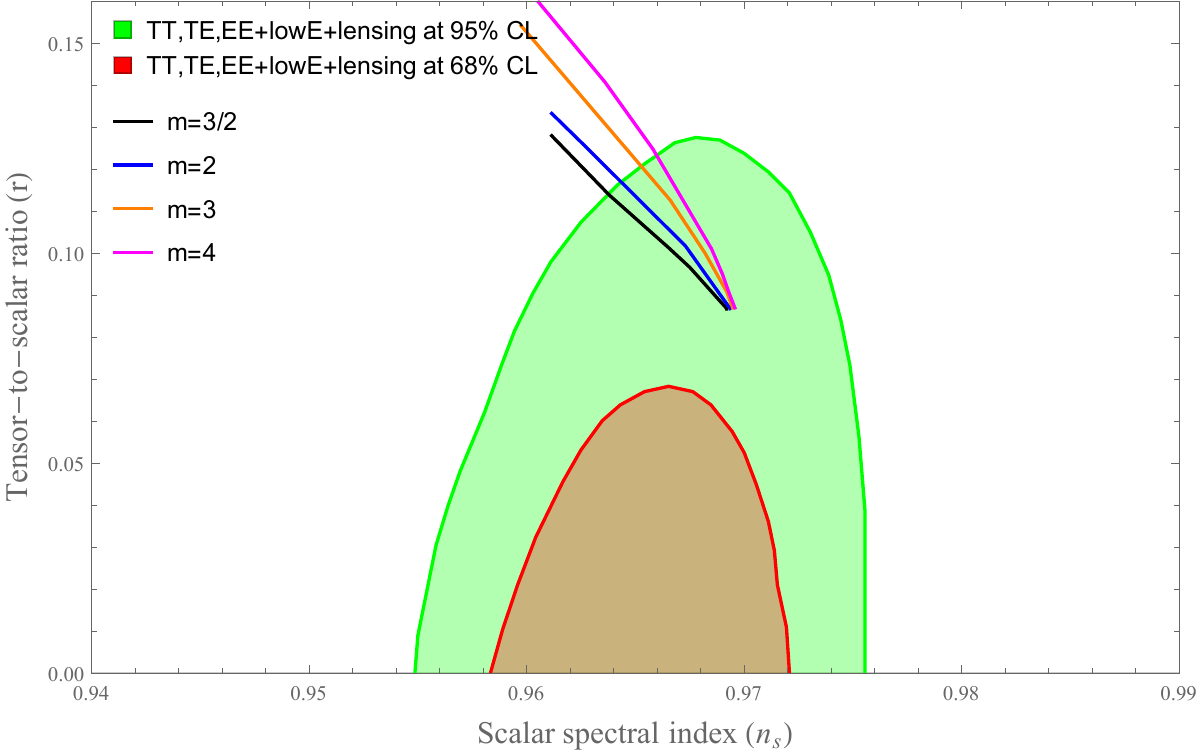}
        \caption {The $n_s-r$ plot predicted by the hilltop potentials corresponding to $m = \{3/2,2,3,4\}$ marked by black, blue, orange and pink colour respectively at $N=50$ and $\mu=5M_p$. The marginalized joint 68\% and 95\% C.L. regions for $n_s$ and $r$ at $k=0.002 Mpc^{-1}$ from Planck 2018 data are shown in red and green colour respectively \cite{r39}.\label{f2}}
\end{figure}

In this case, we pursued the same calculation fixing the $vev$ of the inflaton at $5M_p$ scale. The values of $n_s$ and $r$ are obtained at 50 e-folds and plotted for each cases of $m$ as shown in the Figure \ref{f2}. In the plot we observe similar trend as seen in the earlier case. The lines for $m = \{3/2,2,3,4\}$ fall into the $2\sigma$ region of Planck 2018. This implies hilltop potentials can predict result in conformity with the Planck 2018 data. The increase in positive values of $\lambda$ increases $n_s$ and suppresses $r$. However, $n_s$ tends to be constant at larger values of $\lambda$.   \\
 \\
Finally, the range of the model parameter $\lambda$ is constrained for which the potentials remain consistent with the Planck data. It is presented in the Table \ref{t2}. It is observed that when the scale is lowered, the allowed model parameter space shifts to higher values for each cases of hilltop potential.
\begin{table}[hbt!]
\centering
            \begin{tabular}{lccc}
            \hline \hline
    Case   &   Range of Model Parameter \\ \hline
   $m=3/2$	&	 $6 < \lambda < 79$	\\
   $m=2$	&    $12 < \lambda < 89$	\\
   $m=3$	&    $30 < \lambda < 108$	\\
   $m=4$	&	 $55 < \lambda < 127$	\\
 \hline \hline
            \end{tabular}
             \caption{Allowed parameter space for $m = \{3/2,2,3,4\}$ cases for $N=50$ at $\mu = 5 M_p$ }
            \label{t2}
            \end{table}
\\
It is pertinent to note that $n_s$ and $r$ are sensitive to positive values of $\lambda$ only. For negative values of $\lambda$, this model is unable to realise inflation with hilltop potentials.

\section{Conclusion}
Modified theories of gravity are at the forefront of current theoretical research. It bridges the early and late-time evolution, and simultaneously explains astrophysical compact objects. Having this motivation we started with the aim to study hilltop inflation models within $f(\phi,T)$ gravity, a newly proposed extended version of $f(R,T)$ gravity. We choose the simplest coupling term $f(\phi,T)=\sqrt{\kappa} \lambda \phi T$ in the gravitational Lagrangian and studied its effect on the cosmological observables viz. scalar spectral index $n_s$ and tensor-to-scalar ratio $r$. It is observed that both $n_s$ and $r$ are sensitive to this coupling term. We found that for a certain region of the model parameter space $\lambda$, the different hilltop potentials characterised by $m=\{3/2,2,3,4\}$ agree with the Planck 2018 bound and their respective $n_s - r$ plot enter into the $2\sigma$ region.\\
\\
Another concern with the hilltop potentials is its requirement of super Planckian scales ($\mu \gg 10M_p$) to meet current observational bounds. This issue has been addressed in our study and it is found that the scale $\mu$ can be lowered below $10M_p$ within the $f(\phi,T)$ gravity setup, staying in agreement with Planck 2018 data. We presented results for hilltop potentials with $m=\{3/2,2,3,4\}$ cases at $10M_p$ and $5M_p$ scales in the manuscript. It is observed that as we scale down from $10M_p$ to $5M_p$, the $n_s - r$ plot remains within $2\sigma$ region however the range of $\lambda$ shifts to higher values in all cases. It is important to note here that the scale $\mu$ can further be lowered by fine-tuning $\lambda$ at larger values.\\
\\
In this work, the trace of EM tensor $T$ is calculated from the simple canonical scalar field only. But it can be sourced from other scalar fields as well as from multiple scalar fields. This might present interesting results. Keeping this in mind, non-canonical scalar fields and multiple scalar fields can be incorporated into the action for investigation which we leave as a scope for future work. Further, the coupling term has the functional form $f(\phi,T)$, hence higher order terms of $T$ can also be inserted in the action for future investigation.


\end{document}